\documentclass[aps,prA,twocolumn,superscriptaddress]{revtex4-1}
\usepackage{graphicx}
\usepackage{setspace}
\usepackage{amsmath}
\usepackage{amssymb}
\usepackage{color}
\usepackage{array}
\usepackage{subfigure}
\usepackage{hyperref}
\usepackage{float}
\begin{document}
\title{Quasi-1D scattering with general transverse 2D confinement}
\author{Chen Zhang}
\affiliation{Department of Physics and JILA, University of Colorado, Boulder, Colorado 80309-0440, USA}
\affiliation{Department of Physics, Purdue University, West Lafayette, Indiana 47906, USA}
\author{Chris H. Greene}
\affiliation{Department of Physics, Purdue University, West Lafayette, Indiana 47906, USA}
\date{\today}
\begin{abstract}
Confinement induced resonances (CIR) in quasi-1D systems have been theoretically predicted and observed in various ultracold atomic gases. Here a regularized local frame transformation method is developed to treat CIR in a quasi-1D system that has an arbitrary transverse trap shape in general. The method is applied to predict the CIR position in a system having a square transverse geometry with hard walls. 
\end{abstract}

\maketitle

\section{Introduction}
Quantum systems in reduced dimensions have attracted extensive interest dating back at least to the 1950s. Since then, striking predictions have been made for the behavior of quantum gases in reduced dimensions. In both many-body \cite{gross:195, girardeau:516, Girardeau1965, Haldane1981, Dunjko2001} and few-body physics\cite{dodd:207,mcguire:622, Blume2012}, predictions exist that quantum states in reduced dimensional systems will show qualitatively different behavior from isotropic 3D systems, both in the zero-temperature limit and in the thermodynamic limit. Many of these conclusions can be connected to theoretical explorations from the mathematics community into the nature of scattering in a space of arbitrary dimensions \cite{adawi:358}, dating even farther back in time. However, for a long time, the creation and observation of reduced dimensional systems were hampered by limits on existing experimental techniques. Later, after laser cooling and trapping techniques were developed for ultracold atomic gases \cite{Grimm200095}, many experimental quasi-1D and quasi-2D \cite{Greiner2001, Gorlitz2001, Boyd2007, Bloch2012} quantum systems were realized and a number of theoretically-predicted novel behaviors were observed. In quasi-1D fermionic $^{40}$K \cite{Moritz2003, Gunter2005} and bosonic $^{87}$Rb \cite{Gorlitz2001}, various properties from pure-1D theory have been confirmed. By tuning the ratio(s) of trapping frequencies in different directions ($\omega_{\rho}/\omega_{z}$ \textit{e.g.}), the shape of an ultracold gas can be engineered and varied continuously from a nearly isotropic 3D geometry to a quasi-1D or quasi-2D geometry. Cross-dimensional effects have been observed \cite{Eisenstein2004, Lamporesi2010} and also theoretically studied \cite{Nishida2010, Blume2012}. These phenomena in low-dimensional systems have recently been reviewed by \cite{Manninen2012119, Dunjko2011461, Yurovsky200861, RevModPhys.80.885, RevModPhys.78.1111}.
\par
After achieving success in creating reduced-dimensional systems, physicists began to explore the scattering properties of atoms, hoping to find additional tunability of interactions in ultracold gases. The confinement-induced resonance(CIR) is one such useful tunability of particular interest. From an experimental point of view, the CIR adds another way to manipulate and control low-dimensional systems. Dunjko \textit{et.al} \cite{Dunjko2011461} has reviewed developments related to CIR for a variety of systems. The CIR is unique in the sense that the system exhibits the unitarity limit at a non-resonant value of the 3D scattering parameters (\textit{i.e.} the scattering length for \textit{s}-wave, and the scattering volume for \textit{p}-wave). 
\par
The CIR in a harmonically transverse-confined quasi-1D system was predicted theoretically by Olshanii \cite{Olshanii1998}, and later extended to fermions by Granger and Blume \cite{Blume2004}. In realistic atomic interactions, the incoming wave in the z-direction will not only couple to the \textit{s}-wave (or \textit{p}-wave for fermions) component of the interaction, but also to higher partial waves components. Higher partial wave contributions to the CIR become increasingly important as the scattering energy increases. To address this point, Giannakeas \textit{et.al} \cite{Giannakeas2012} recently developed a method involving all partial waves, and pointed out a correction to the position of the CIR that derives from the \textit{d}-wave component. 
\par
In the treatment of $s$-wave scattering, Olshanii \cite{Olshanii1998} derived the exact wave function and phase shift for a regularized zero range potential that are associated with low energy \textit{s}-wave scattering. This problem can also be treated using Green's function methods. Implementation of an eigenfunction expansion of the Green's function for treating a bound state induced by a s-wave zero-range potential was discussed in \cite{Englert1998}. A similar eigenfunction expansion of Green's functions is embedded in Olshanii's original work, although not explicitly emphasized. Later, a systematic discussion of the Green's function method and the corresponding Lippman-Schwinger equation was presented in \cite{PhysRevLett.93.170403, Naidon2007}. Recently the Green's function method was applied to treat $s$-wave zero-range interaction in asymmetric transverse harmonic confinement \cite{PhysRevA.83.053615}. However for higher partial waves, different mathematical models of the zero-range interaction are needed in the Olshanii treatment \cite{Idziaszek2006}. In an alternative treatment, Granger and Blume \cite{Blume2004} used the frame transformation method, in which the 3D phase shift information directly determines the 1D reaction matrix $K$. Thus the frame transformation approach avoids the need to design a zero-range model potential for higher partial wave scattering, which has some conceptual advantages. For example, a $p$-wave CIR was predicted, and later observed experimentally \cite{delCampo2006}. 
\par
As the exeperimental techniques of laser trapping have grown in sophistication, various confinement potentials beyond the harmonic trap have been realized, \textit{e.g.} optical lattice traps \cite{Haller04092009}, uniform trap \cite{PhysRevLett.110.200406}. In these systems, a more general theory of CIR positions beyond a harmonic trap is needed. Kim \textit{et. al} gave a general description of a symmetric cylindrical hard wall trap \cite{PhysRevA.72.042711} using a Green\rq{}s function method. The aim of the present work is to develop a systematic description of quasi-1D scattering in arbitrary transverse confinement. Specifically, we consider the situation in which one light particle is scattered off an infinitely massive particle in the center of the confinement. We apply the local frame transformation method \cite{Fano1981}, and solve the divergence problem occurring in this method. Because this divergence also arises in many other ananlytical and numerical applications of this method, we believe that the systematic discussion presented here yields valuable insights. To illustrate our implementation of the method, we treat the CIR in a square well transversely-confined system as an example. The development of our regularization method is inspired by the s-wave regularized delta function in 3D, and it can be generalized to higher partial waves interacting through short-range potentials. 
\section{Regularized frame transformation}
A physical system can often be divided into different regions, where the controlling physics varies. In each region, because of the difference in the symmetry of the interactions or the boundary conditions, the good quantum numbers can differ from one region to another. Specifically, the quasi-1D system can be divided into two region: $r\ll L_{\perp}$ and $|z|\gg L_{\perp}$, and the characteristic length $L_{\perp}$ in the transverse confinement is the relevant length scale to divide this system (it is the oscillator length in the case of harmonic confinement, and the box side length for the square well confinement), $r$ is the distance from the scattering center to the light particle, $z$ is the distance in z-direction, which is the only asymptotic outgoing direction. At large distance ($|z| \gg L_{\perp}$) in any particular open channel ($E=\dfrac{\hbar^2k^2}{2\mu}> E_{n_x, n_y}$ defines the term ``open channel" with respect to its channel energy $E_{n_x, n_y}$, in which $(n_x, n_y) are index of the eigenstates (channel) in the transverse plane$), the outgoing wave vector $k_z$ is an asymptotically good quantum number. Thus the phase shift for each $k_z$ asymptotically approaches to a constant. However, at very short range near the center of the confinement, where a scattering event occurs, the interaction potential has the full 3D spherical symmetry. In order to connect the short range scattering event and the large distance observables, we need to project the scattering information near the spherically symmetric scattering center onto the transverse channels to yield quasi-1D scattering information in the z-direction. The determination of this projection is the main goal of the local frame transformation method. The term ``local" in this context was originally coined by Fano \cite{Fano1981} to distinguish this method from the usual unitary class of frame transformations in \cite{PhysRevLett.55.1066,PhysRevA.57.2407}. Later, this method was extended and applied to several scattering problems \cite{Fano1981, Harmin1982, WongGreene1988, Greene1987}, and in many cases it has resulted in excellent agreement with experimental observations and higher efficiency than previous theories. Note that whereas Zhao {\it et al.} \cite{PhysRevA.86.053413} claim to have identified a flaw in the Fano technique for transformation of the solution that is irregular at the origin, the present application is tested numerically in Sec.III below and it exhibits no sign of any such difficulty.
\par
This section is organized as follows, first the of CIR is described for general 2D confinement; then some comparisons are made with the well-studied CIR in 2D iostropic harmonic confinement; next our method is applied to a transverse square cross section confinement with hard walls, which is one of the simplist geometries for 2D confinement without azimuthal symmetry.
\subsection{Quasi-1D CIR in General 2D confinement}
In this paper, we specifically address the problem of one light particle (mass $\mu$) that scatters from an infinitely massive particle located at the trap center. This is a system that could be realized experimentally, though apparently it has not yet been demonstrated. For this system it is not necessary to condier any coupling between the relative and center of amss coordinates of the colliding pair of particles. The Hamiltonian for the light particle considered in this problem is,
\begin{equation}
H=H_{\perp}(x,y)-\dfrac{\hbar^2}{2\mu}\dfrac{\partial^2}{\partial z^2}+V_{\mbox{int}}(x,y,z).
\end{equation}
The frame transformation method can be applied to this problem, because the Hamiltonian has two clear limits: $r\ll L_{\perp}$ and $|z|\gg L_{\perp}$. $H_{\perp}$ is the ``transverse'' part of the Hamiltonian, and its eigenvalues define the ``channel threshold energies". $V_{\mbox{int}}$ is the short range ($r_c$ is its range) spherically symmetric potential, which is not separable in the Cartesian coordinates where the boundary conditions are separable. In the remainder of this article, unless otherwise stated, a system of units is adopted for which $\hbar=1$ and $\mu=1$.
\par
This model Hamiltonian is exact for the problem of one light particle that scatters from an infinitely massive particle located at the trap center. When it comes to treating two interacting particles with short range interaction, having both finite massesthis treatment might still be a reasonable approximation, or at least a first step in a more general theory that also accounts for nonseparability of the relative and center-of-mass (COM) degrees of freedom. Some discussion of the non-separability of COM and relative motion are presented by Sala \textit{et. al} \cite{PhysRevLett.110.203202} in the context of optical lattices, for which the anharmonicity is weak.
\par
The eigenfunctions at short distance $r_c\ll r\ll L_{\perp}$ can be represented as linear combinations of non-interacting channel functions:
\begin{equation} 
\Psi_{lm\epsilon}(\mathbf{r})=\sum_{l^{\prime}m^{\prime}}F_{l^{\prime}m^{\prime}\epsilon}\delta_{l,l^{\prime}}\delta_{m, m^{\prime}}-G_{l^{\prime}m^{\prime}\epsilon}K_{l^{\prime}m^{\prime},lm}^{3D}(\epsilon),
\end{equation}
where the reaction matrix $\underline{K}^{3D}$ encapsulates all the information about the short-range scattering caused by the potential, $l$ is the orbital angular momentum quantum number, \textit{m} is the orbital magnetic quantum number. $F$ and $G$ are the energy-normalized regular and irregular solutions in spherical coordinates for a given scattering energy $\epsilon$:
\begin{equation}
\begin{cases}
F_{lm\epsilon}(\mathbf{r})=Y_{lm}(\hat{\mathbf{r}})f_{l\epsilon}(r) & f=\sqrt{\dfrac{2k}{\pi }}j_{l}(kr)\\
G_{lm\epsilon}(\mathbf{r})=Y_{lm}(\hat{\mathbf{r}})g_{l\epsilon}(r) & g=\sqrt{\dfrac{2k}{\pi }}n_{l}(kr),\\
\end{cases}
\end{equation}
$f$ and $g$ are the energy-normalized spherical Bessel functions, $f$ is regular at $r\to 0$, $g$ is irregular at $r\to 0$, and they oscillate as $r\to \infty$ with a $\pi/2$ difference in phase. The $z$-parity quantum number of the solutions is $(-1)^{l-m}$, which is important to keep in mind, since later the 3D solutions will be projected onto outgoing waves in the $z$-direction. The 2D isotropic harmonic confinement is a special case, in which the $m$ quantum number is conserved in both the spherical and cylindrical symmetries, so there is no need to explicitly specify $m$ \cite{Blume2004}. However, for a 2D transverse trap in general, all $l-$ and $m-$values could be coupled. In the unitary limit, of course, with short-range interactions, the lowest one or two partial waves for each symmetry are expected to dominate the scattering observables.
\par
The regular $\psi$ and irregular $\chi$ eigenfunctions at large distances $L_{\perp}\ll |z|$ are:
\begin{equation}
\begin{cases}
\psi_{n_x,n_y,q}(\mathbf{r})=\\
\\
\phi_{n_x,n_y}(x,y)(\pi q)^{-\frac{1}{2}}
\begin{cases}
\cos(qz), &\pi_z=+1 \\
\sin(qz), &\pi_z=-1
\end{cases}
 \\
\chi_{n_x,n_y,q}(\mathbf{r})=\\
\\
\phi_{n_x,n_y}(x,y)(\pi q)^{-\frac{1}{2}}
\begin{cases}
\sin(q|z|), &\pi_z=+1\\
-sign(z)\cos(qz), & \pi_z=-1
\end{cases}
\end{cases}
\end{equation}
where $n_x$, $n_y$ are quantum numbers in the transverse plane. For example, in the square well cross section case, the eigenstates in the transverse plane for the even parity, of immediate interest here, are $\frac{2}{L}\cos(\frac{(2n_x+1)\pi}{L_{\perp}}x)\cos(\frac{(2n_y+1)\pi}{L_{\perp}}y)$. The confinement in the $x-y$ plane forces the scattered waves to be asymptotically directed along either the positive or negative $z$-axis.
\par
The evolution of the spherical wave function near the scattering center ($r_c\le r\ll L_{\perp}$) into an asymptotic wave function at $|z|\gg L_{\perp}$ on the a fixed energy shell $\epsilon$ can be accomplished by projecting the eigenfunctions near the scattering center onto the eigenfunctions relevant in the asymptotic region:
\begin{equation}
\begin{array}{lll}
\psi_{n_x,n_y,q_{n_x,n_y}}(\mathbf{r})=\sum_{l,m} U_{n_x,n_y,q_{n_x,n_y};l,m,\epsilon}F_{lm\epsilon}(\mathbf{r}),\\
\\
\chi_{n_x,n_y,q_{n_x,n_y}}(\mathbf{r})=\sum_{l,m} {(U^{-1})^{T}}_{n_x, n_y,q_{n_x,n_y};l,m,\epsilon}G_{lm\epsilon}(\mathbf{r}),
\end{array}
\end{equation}
$\underline{U}$ is the transformation matrix between the two sets of eigenstates, and the $\underline{U}$ matrix is energy dependent, since the expansion is performed on the energy shell. The summation includes only solutions of the same $z$-parity. Two examples of $\underline{U}$ are shown in appendix.

As was shown in previous papers \cite{Fano1981, Harmin1982}, the local frame transformation is non-unitary. This is because the two sets of eigenstates solve two different partial differential equations (actually, in the present case, they solve the PDE with the same Hamiltonian but different boundary conditions), which are only approximately equal to each other in the region $r_c \le r \ll L_{\perp}$ \cite{Fano1981}. Subsequently a number of studies \cite{Greene1980, Greene1987, Blume2004, Giannakeas2012} have discussed the resulting transformation of the reaction matrix $K$, namely: $\underline{K}^{1D}=\{K_{n_1n_2,{n_1}^{\prime}{n_2}^{\prime}}\}=\underline{U}^{T}\underline{K}^{3D}\underline{U}$.
The physical $K^{1D}$ matrix is deduced from closed-channel elimination \cite{Greene1987}, which enforces exponential decay in all energetically closed channels at $|z|\to \infty$:
$\underline{K}_{oo}^{1D,phys}=\underline{K}_{oo}^{1D}+i\underline{K}_{oc}^{1D}(1-i\underline{K}_{cc}^{1D})^{-1}\underline{K}^{1D}_{co}$,
in which $c$ denotes closed channels, $o$ denotes open channels. All the information from closed channels is encapsulated into $\underline{K}^{1D,Phys}$. When one channel is open for each overall symmetry (even or odd $z$-parity), the poles of the physical reaction matrix $\underline{K}^{1D, Phys}$ predict the positions of each CIR, which are values of the scattering length and/or trapping geometry for which $Det[1-iKcc]\sim 0$ at vanishing incident kinetic energy in the lowest channel. For the case of only one open channel, the lone physical $K$ matrix element is the tangent of the quasi-1D scattering phase shift. This quasi-1D scattering phase shift in the single open channel recovers the pure-1D feature in the low energy limit (since there is no concept of ``closed channels" in a pure-1D system), having the form $\tan\delta=-\dfrac{g_{1D}}{k}$. Clark pointed out this fact in \cite{Clark1983}, which is equivalent to an observation that a ``sudden" $\pi/2$ change occurs in the low energy scattering phaseshift when just an infinitesimal $g_{1D}$ is turned on. We illustrate this point in Fig. \ref{Quasi1Ddelta}.
\par
In addition to all the general properties mentioned above concerning quasi-1D scattering, the properties of the short-range potential can in some cases allow a further simplification of the problem. If $\underline{K}^{3D}$ is block-diagonal and only one angular momentum partial wave dominates the 3D scattering phase shift, a good approximation to the $K$ matrix is $K^{3D}_{lm,l^{\prime}m^{\prime}}=\delta_{l,l^{\prime}}\delta_{m,m^{\prime}}\tan(\delta_{l})$, $\underline{K}_{oo}^{1D,Phys}(E)=\underline{K}_{oo}^{1D}[1-i\lambda_c(E)]^{-1}$, where $\lambda_c=Tr\underline{K}_{cc}^{1D}=\tan\delta_{l}\sum_{\mbox{closed channels}}(U_{n_x,n_y,q_{n_x,n_y};l,m,\epsilon})^2$ \cite{Blume2004}. Under this single partial wave approximation, each pole of the physical $\underline{K}^{1D,Phys}$ is determined by the vanishing of $1-i\lambda_c(E)$.
\par
In general, the summation of squared transformation matrix elements can be expressed as a summation over all closed channel quantum numbers ($n_x$, $n_y$ as mentioned above) in the 2D confinement portion of Hilbert space. Under the single partial wave approximation, 
\begin{equation}
\begin{array}{lll}
\lambda_c=\tan\delta_{l}\displaystyle\sum^{\prime}(U_{n_x,n_y,q; l,m,\epsilon})^2\\
\\
=\tan\delta_{l}\displaystyle\sum^{\prime}\dfrac{1}{kD_{\perp}}(\epsilon_{n_x,n_y}-\epsilon)^{\frac{1}{2}},
\end{array}
\end{equation}
in which $D_{\perp}$ is one characteristic length of the transverse confinement, $\epsilon_{n_x,n_y}$ is the channel energy divided by the characteristic energy in the system (usually associtated with the characteristic length), $\epsilon$ is the channel energy divided the same characteristic energy, $\displaystyle\sum^{\prime}$ indicates the summation over all $(n_x, n_y)\in\mbox{closed channels}$.
\par
\subsection{Examples for various types of confinement}
\subsubsection{2D isotropic harmonic confinement}
In the treatments of this quantity $\lambda_c(E)$ in 2D isotropic harmonic confinement \cite{Blume2004,Giannakeas2012}, its expression takes one of the following forms as an infinite summation:
\begin{equation}
\begin{array}{lll}
\lambda_{c}^{l=1}=\dfrac{V_{p}}{a_{\perp}^3}\displaystyle\sum_{n=1}^{\infty}\sqrt{n+\dfrac{3}{2}-\epsilon} \mbox{  ($p$-wave)}\\
\\
\lambda_{c}^{l=0}=\dfrac{a_s}{a_{\perp}}\displaystyle\sum_{n=1}^{\infty}\dfrac{1}{\sqrt{n+\dfrac{3}{2}-\epsilon}} \mbox{  ($s$-wave)}.
\end{array}
\end{equation}
Moreover, in the $s$-wave case, Olshanii \cite{Olshanii1998} encountered this same summation of squared transformation matrix elements, which is clearly divergent. A regularization method is developed in the following paragraphs, motivated by other analyses that have regularized different types of zeta functions in various contexts.
\par
The relevant special function for the 2D isotropic harmonic oscillator confinement is the Hurwitz zeta function. Observe that in the definition of the Hurwitz zeta function: 
\begin{equation}
\zeta_{H}(s,q)=\sum_{n=0}^{\infty}\dfrac{1}{(q+n)^s}, (s>1)
\end{equation}
the same functional form is present in the region $s\in (1,\infty)$ where the series is convergent. So the analytical continuation of the Hurwitz zeta function from $s\in (1, \infty)$ to $s\in (-\infty, 1]$ can potentially regularize the divergent summation and yield a physically relevant value. A similar spirit has been implemented in the calculation of the Casmir force between two infinitely large planes and between other shapes of conductors \cite{Plunien198687, Javier2003}. The extensive work on the Hurwitz zeta function in the mathematics community suggests a way to develop the regularization in general. For the present quasi-1D scattering system, we implement a regularization procedure that has been utilized in the mathematics community. The summation can be viewed as the following limit process:
\begin{equation}
\Lambda_H(\xi,\epsilon)=\dfrac{\partial}{\partial \xi}\left.\left(\xi\sum_{n=0}^{\infty}\dfrac{\exp(-\xi\sqrt{n+1+\epsilon})}{\sqrt{n+1+\epsilon}}\right)\right|_{\xi \to0}.
\label{summationHO}
\end{equation}
This summation is not uniformly convergent, which means that interchanging the order of taking the limit $\xi\to 0$ and summing over all terms might not be justified. For any finite $\xi$, the convergence of the summation is guaranteed by the exponential suppression at large $n$. For $\xi=0$, however, the summation assumes the same form encountered in the $l=0$ case. The divergence in the local frame transformation approach can be viewed as having originated in an unjustifiable interchange of the order of these two operations. 
\par
Olshanii \textit{et.al} \cite{Olshanii1998, Dunjko2011461} encountered the same divergence problem while solving for the scattering amplitude in the open channel. They treated the divergence as follows. The summation in Eq. \ref{summationHO} can be expanded as a power series in $\xi$ (where $\xi=\frac{z}{L_{\perp}}$) near the energy of the lowest threshold, giving: 
\begin{equation}
\Lambda_H(\xi,\epsilon)=F(\xi)+\tilde{\Lambda}_H(\xi,\epsilon).
\end{equation}
Here $F$ is the integral approximation of the summation right at threshold, $F(\xi)=\int_{0}^{\infty}d\nu\frac{\exp(-\xi\sqrt{\nu})}{\sqrt{\nu}}=\frac{2}{\xi}=\sum_{s=1}^{\infty}\int_{s-1}^{s}d \nu \frac{\exp(-x\sqrt{\nu})}{\sqrt{\nu}}$, which exhibits the divergence with respect to $\xi$. $\tilde{\Lambda}_{H}$ is the regular, physically relevant part of $\Lambda_{H}$.  At an arbitrary scattering energy, the integration approximation becomes $\int_{1}^{N}d\nu \frac{\exp(-\xi\sqrt{\nu+\epsilon})}{\sqrt{\nu+\epsilon}}=\frac{2}{\xi}(\exp(-\xi\sqrt{1+\epsilon})-\exp(-\xi\sqrt{N+\epsilon}))$. 
Thus the summation in Eq.\ref{summationHO} can be cast as:
\begin{equation}
\begin{array}{lll}
\Lambda_H(\xi,\epsilon)=\dfrac{2}{\xi}+\tilde{\Lambda}_H(\xi,\epsilon)\\
\\
=\dfrac{2}{\xi}-\dfrac{2}{\xi}+2\dfrac{\exp(-\xi\sqrt{1+\epsilon})}{\xi}+\\
\\
\lim_{N\to\infty}(-\dfrac{2}{\xi}(\exp(-\xi\sqrt{1+\epsilon})-\exp(-\xi\sqrt{N+\epsilon}))+\\
\\
\sum_{n=1}^{N}\dfrac{\exp(-\xi\sqrt{n+\epsilon})}{\sqrt{n+\epsilon}})\\
\\
=\dfrac{L_{-1}(\epsilon)}{\xi}+\tilde{\Lambda}_H(\xi\to0^{+},\epsilon)+\mbox{terms that vanish at $\xi=0$},
\end{array}
\label{LambdaHexpansion}
\end{equation}
in which:
\begin{equation}
\tilde{\Lambda}_H(\xi\to0^{+},\epsilon)=\lim_{N\to\infty}\sum_{n=1}^{N}\dfrac{1}{\sqrt{n+\epsilon}}-2\sqrt{N+\epsilon}.
\end{equation}
The present study applies a more general concept to the divergent summation that has been developed in the mathematics community. Our treatment does not rely on the \textit{s}-wave nature of the scattering, and can be directly deduced from the frame transformation result. The following illustration of our method is developed specifically for the square transverse trap geometry with hard walls, because there are two quantum numbers in a general separable 2D confinement, and the 2D square well trap is a more general example with quite different characteristics than geometries for which this type of analysis has previously been worked out. In addition, we show that the isotropic 2D harmonic oscillator geometry ($\omega_x=\omega_y=\omega_{\rho}$) is a special case in our treatment.
\par
\subsubsection{Transverse 2D square confinement geometry}
\begin{figure}
\includegraphics[scale=0.18]{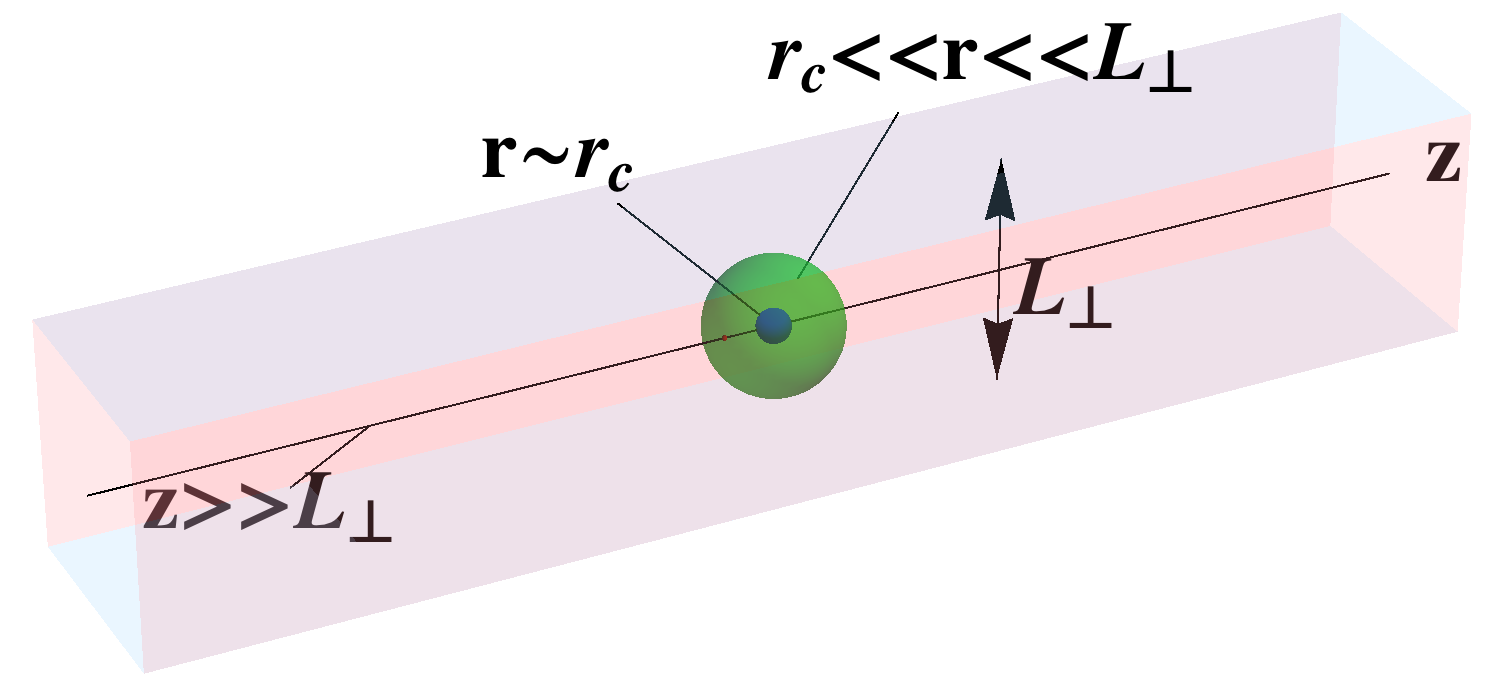}
\caption{This is a sketch of the 2D square well confinement geometry, and the separation of differen length scales in the quasi-1D scattering process.}
\label{SquareWellFig}
\end{figure}

The geometry of the transverse 2D square hard wall confinement is depicted in Fig. \ref{SquareWellFig}.
The summation of squared $\underline{U}$-matrix elements for the 2D square confinement geometry is: 
\begin{equation}
\displaystyle\sum_{n_x,n_y \ge 0}^{(n_x,n_y)\ne (0,0)} \dfrac{1}{\sqrt{(n_x+\dfrac{1}{2})^2+(n_y+\dfrac{1}{2})^2+\epsilon}}.
\end{equation}
This sum, which must be evaluated before differentiation, takes the following form:
\begin{equation}
\begin{array}{lll}
\Lambda_{E}(\xi,\epsilon)=\dfrac{\partial}{\partial \xi}\left(\xi\displaystyle\sum^{\prime} \dfrac{\exp(-\xi\sqrt{(n_x+\dfrac{1}{2})^2+(n_y+\dfrac{1}{2})^2+\epsilon})}{\sqrt{(n_x+\dfrac{1}{2})^2+(n_y+\dfrac{1}{2})^2+\epsilon}}\right)|_{\xi\to0^{+}}\
\\
=\dfrac{1}{4}\dfrac{\partial}{\partial \xi}\left(\xi\displaystyle\sum_{n_x,n_y}\dfrac{\exp(-\xi\sqrt{(n_x+\dfrac{1}{2})^2+(n_y+\dfrac{1}{2})^2+\epsilon})}{\sqrt{(n_x+\dfrac{1}{2})^2+(n_y+\dfrac{1}{2})^2+\epsilon}}\right)_{\xi\to0^{+}}\\
\\
-\dfrac{\exp(-\xi\sqrt{\frac{1}{2}+\epsilon})}{\sqrt{\frac{1}{2}+\epsilon}}.
\end{array}
\end{equation}
The summation $\displaystyle\sum^{\prime}$ is over all closed-channel quantum numbers in one quarter of the 2D $(n_x,n_y)$ plane. 
\par
\begin{figure}
\includegraphics[scale=0.48]{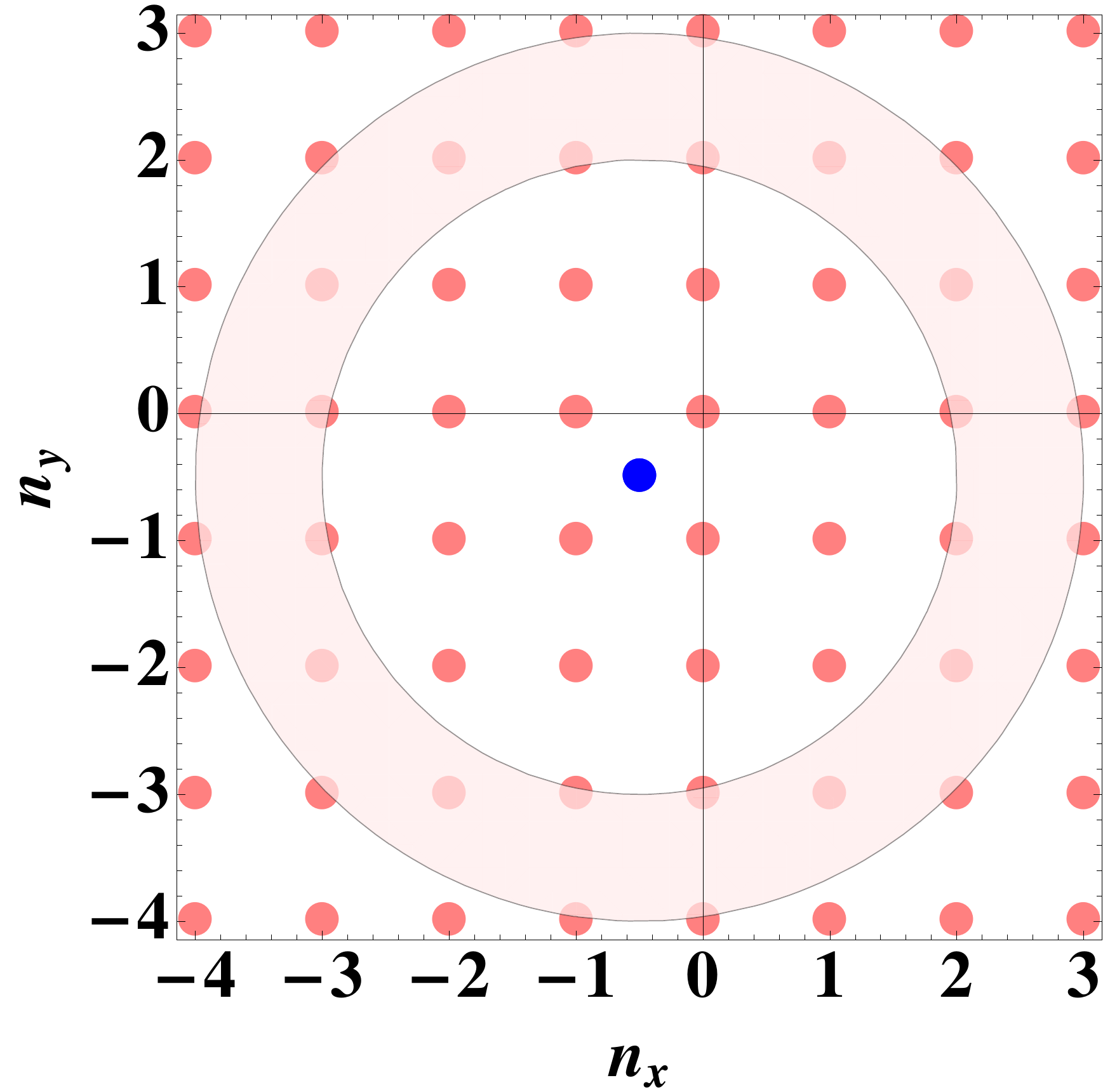}
\caption{Plot of quantum numbers $\mathbf{n}=(n_x,n_y)$ that are summed over in the 2D integer plane, shown as smaller circles (pink online). The ring region is defined by the range $(2n-1)^2/2\le |\mathbf{n}+(\frac{1}{2},\frac{1}{2})|^2\le\frac{(2n+1)^2}{2}$, with $n=3$. The large point (blue online) is at $(-\frac{1}{2},-\frac{1}{2})$. }
\label{2DIntegerPlot}
\end{figure}

The number of points in one ring $\frac{1}{2}(2n-1)^2<|\mathbf{n}+(\frac{1}{2},\frac{1}{2})|^2\le\frac{1}{2}(2n+1)^2$ can be understood as the density of states in the quantum number space, $\mathbf{n}=(n_x,n_y)$. The summation over all the quantum numbers can be approximated by an integral over the density of states in the 2D quantum number space, and this approximation can be used to separate out the nature of the infinite sum singularity. The integral can be evaluated by changing to the polar plane:
\begin{equation}
\begin{array}{lll}
\int_{-\infty}^{\infty}\int_{-\infty}^{\infty} d n_x d n_y \dfrac{\exp(-\xi\sqrt{(n_x+\frac{1}{2})^2+(n_y+\frac{1}{2})^2+\epsilon})}{\sqrt{(n_x+\frac{1}{2})^2+(n_y+\frac{1}{2})^2+\epsilon}}\\
\\
=\lim_{N\to\infty}\displaystyle{\int}_{\frac{\sqrt{2}}{2}}^{\frac{(2N+1)\sqrt{2}}{2}}2\pi r dr \dfrac{\exp(-\xi\sqrt{r^2+\epsilon})}{\sqrt{r^2+\epsilon}}\\
\\
=\lim_{N\to\infty}\sum_{n=1}^{N}\displaystyle{\int}_{(2n-1)\sqrt{2}/2}^{(2n+1)\sqrt{2}/2}2\pi r dr \dfrac{\exp(-\xi\sqrt{r^2+\epsilon})}{\sqrt{r^2+\epsilon}}\\
\\
=\lim_{N\to\infty}\dfrac{2\pi}{\xi}(\exp(-\xi\sqrt{\frac{1}{2}+\epsilon})-\exp(-\xi\sqrt{\frac{(2N+1)^2}{2}+\epsilon}))
\end{array}
\end{equation}
By using the series expansion of $\Lambda_{E}$ in terms of $\xi$, which has the same spirit of Eq. \ref{LambdaHexpansion}, we have:
\begin{equation}
\begin{array}{lll}
4\Lambda_E(\xi,\epsilon)=\dfrac{2\pi}{\xi}+(-\dfrac{2\pi}{\xi}+\dfrac{2\pi}{\xi}\exp(-\xi\sqrt{\frac{1}{2}+\epsilon}))\\
\\
+\lim_{N\to\infty}\left[ (-\dfrac{2\pi}{\xi}(e^{-\xi\sqrt{\frac{1}{2}+\epsilon}}-e^{\xi\sqrt{\frac{(2N+1)^2}{2}+\epsilon}})).\right.\\
\\
+\left. \sum_{n=1}^{N}\sum_{\frac{1}{2}<|\mathbf{n}+(\frac{1}{2},\frac{1}{2})|^2\le\frac{(2n+1)^2}{2}} \right.\\
\\
\left. \dfrac{\exp(-\xi\sqrt{(n_x+\frac{1}{2})^2+(n_y+\frac{1}{2})^2+\epsilon})}{\sqrt{(n_x+\frac{1}{2})^2+(n_y+\frac{1}{2})^2+\epsilon}})\right]\\
\\
\end{array}
\end{equation}
The terms inside the limiting process $\lim_{N\to \infty}$ can be rearranged as two parts, one of which is 
constant as $\xi \to 0$, while the other vanishes as $\xi \to 0$.
\begin{equation}
4\Lambda_{E}(\xi,\epsilon)=\dfrac{2\pi}{\xi}+4\tilde{\Lambda}_E(\xi\to0^{+},\epsilon)+\mbox{terms vanish at $\xi\to0^{+}$}
\end{equation}
At this point, the form of the $\Lambda_E$ function resembles that of the $\Lambda_H$ function which arises for the harmonic oscillator trap. The corresponding residual part is the regularized summation:
\begin{equation}
\begin{array}{lll}
4\tilde{\Lambda}_E(\xi\to 0^{+},\epsilon)=\sum_{n=1}^{N}\sum_{\frac{1}{2}<|\mathbf{n}+(\frac{1}{2},\frac{1}{2})|^2\le\frac{(2n+1)^2}{2}}\\
\\
\dfrac{1}{\sqrt{(n_x+\dfrac{1}{2})^2+(n_y+\dfrac{1}{2})^2+\epsilon}}-2\pi\sqrt{\frac{(2N+1)^2}{2}+\epsilon}.
\end{array}
\end{equation}
The ring region in $\{n_x, n_y\}$ is demonstrated in Fig. \ref{2DIntegerPlot}. The definition of the term $\tilde{\Lambda}_E(\xi\to0^{+},\epsilon=0)$ is unique, because if one adds a term proportional to $\xi$, it will vanish as $\xi\to 0^{+}$, while the other term $2\pi/\xi$ is the integral approximation to the summation at the threshold energy ($\epsilon=\frac{1}{2}$ in this case), which also has a unique definition.
For 2D harmonic confinement, by the way of contrast, the summation over magnetic quantum numbers is trivially carried out because of the symmetry of that problem.
This analysis calculates the position of the CIR in this geometry to be:
\begin{equation}
\dfrac{a_{s}(\epsilon)}{L_{\perp}}=\dfrac{1}{4\Lambda_E(\xi\to 0,\epsilon)}.
\end{equation}
$\Lambda_E(\xi\to 0,\epsilon)$ turns out to be a special case of the Epstein zeta function \cite{e1994zeta} (chapter $1$, section $2.2$, also briefly discussed below in the Appendix), $a_{s}(\epsilon)$ is the energy dependent 3D scattering length. Many applications of the Epstein zeta function in physics are summarized in \cite{1206.00039}, including its application in the zeta regularization method of high energy physics \cite{Elizalde:2012zza, Hawking1977}.
\par
To test our proposed regularization method, we have first applied it to the Hurwitz zeta function summation. This test has verified that it gives the same numerical value for every tested value of $q$ and $s$ in agreement with other definitions \cite{abramowitz1964handbook, citeulike:4561567} of the Hurwitz zeta function, analytically continued into the region $s\in (-\infty, 1]$.
\section{Numerical calculation of the low energy 1D scattering phase shift}
Our predicted position of the CIR can be tested further by performing a variational $R$-matrix scattering calculation for the square well 2D trap system using a set of B-spline basis functions. The CIR position for a zero-range 3D interaction is extracted by taking its limiting value from finite range model potential calculations. Moreover, the leading finite range correction to the CIR position is also extrapolated and compared with the prediction of effective range theory, in which the scattering length and the effective range of a finite range model potential is defined in terms of small energy expansion of the $s$-wave scattering phase shift $-\dfrac{1}{a_s(E)}=-\dfrac{1}{a_s(0)}+\dfrac{k^2}{2}r_{eff}$. 
\par
\begin{figure}
\includegraphics[scale=0.5]{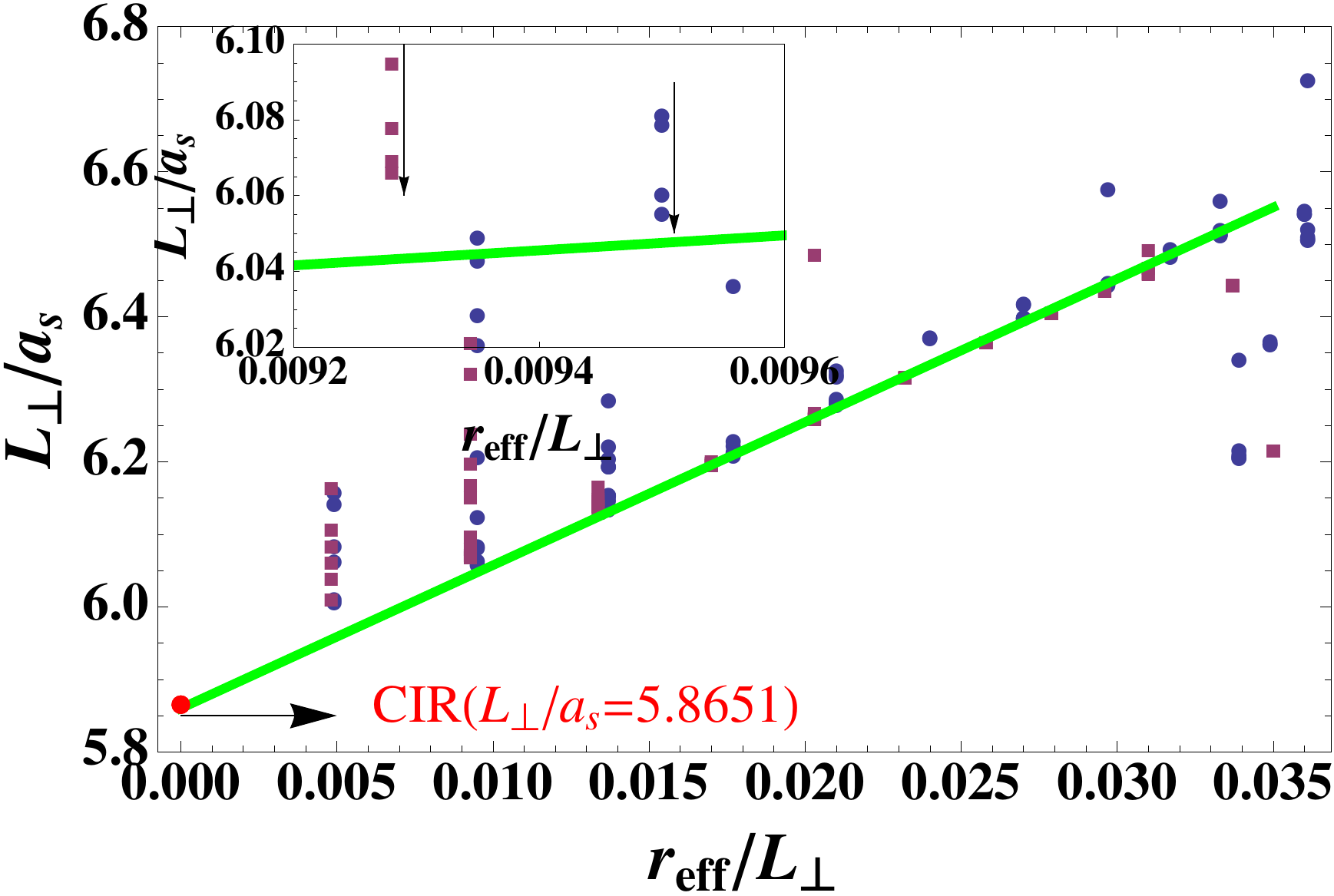}
\caption{The CIR position obtained from finite range interaction model calculations is plotted along with the extrapolation to zero range. The square dots (purple online) are from the potential $V=V_{0}\exp(-\dfrac{r^2}{2d_{0}^2})$, while the circle dots (blue online) are from the potential $V=V_{0}\dfrac{1}{\cosh(\frac{r}{d_{0}})^2}$. The $x$-axis is the effective range of the potential at the position of the CIR. The solid line (green online) is derived from the effective range expansion of the 3D scattering length $|\frac{L_{\perp}}{a_{s}}|=|\frac{L_{\perp}}{a_s(0)}|+L_{\perp}^2 E_{0}\frac{r_{eff}}{L_{\perp}}$, $E_{0}$ is the lowest threshold energy.The single point at $r_{eff}=0$ (red online) is the predicted CIR position from our regularized frame transformation method. The inset shows the pattern of convergence to the analytical value, which is seen in the numerical calculation near $\frac{r_{eff}}{L_{\perp}}\sim 0.010$ (the arrows indicate the better converged numerical values of the CIR position). At larger values of the 3D effective range, the numerical calculation begins to deviate from the analytical prediction. This is because the range of interaction $d_{0}$ in those calculations has become comparable to the CIR scattering length, and hence the ``short" range approximation of the model potential has become less accurate.}
\label{Plot1}
\end{figure}
For short range interactions $d_{0}\ll a_{s}$(CIR), the numerical model calculations agree accurately with the analytical prediction, as is shown in Fig. \ref{Plot1}. The correction plotted versus the effective range of the potential exhibits a linear behavior, and the slope agrees with the prediction from a finite range expansion of the energy dependent scattering length. Some points deviate from the linear relation, as in Fig. \ref{Plot1}. The reason for this is that the width $d_{0}$ of the interaction is comparable to or larger than the CIR scattering length, which sets a new characteristic length scale that limits the applicability of the ``short" range model potential approximation: $\frac{d_{0}}{a_{s}(CIR)} \ll 1$. 
\par
The dot (red online) plotted at $r_{eff}=0$ in Fig. \ref{Plot1} is the prediction from the present regularized frame transformation method, and it agrees with an extrapolation of the numerical calculation to a zero-range potential. The intercept, which corresponds to the resonance position at zero energy, occur in Fig. \ref{Plot1} occurs at $\dfrac{L_{\perp}}{a_{s}(0)}=5.850(\pm0.005)$ from extrapolation of the numerical results, and at $5.864(\pm0.008)$ from the present regularized analytical summation. The slope of the extrapolation line can be deduced from the low-energy effective range expansion of the $s$-wave scattering length: $-\frac{L_{\perp}}{a_{s}(E)}=-\frac{L_{\perp}}{a_{s}(0)}+L_{\perp}^2\frac{k^2}{2}\frac{r_{eff}}{L_{\perp}}$, in which $L_{\perp}$ is the width of the square well, and $\frac{k^2}{2}=E$ is the first threshold energy of the 2D confinement, $a_{s}(E)$ is the energy dependent scattering length, $r_{eff}$ is the effective range of the potential, which usually changes very slowly with $a_{s}(0)$.
\par
\begin{figure}
\includegraphics[scale=0.28]{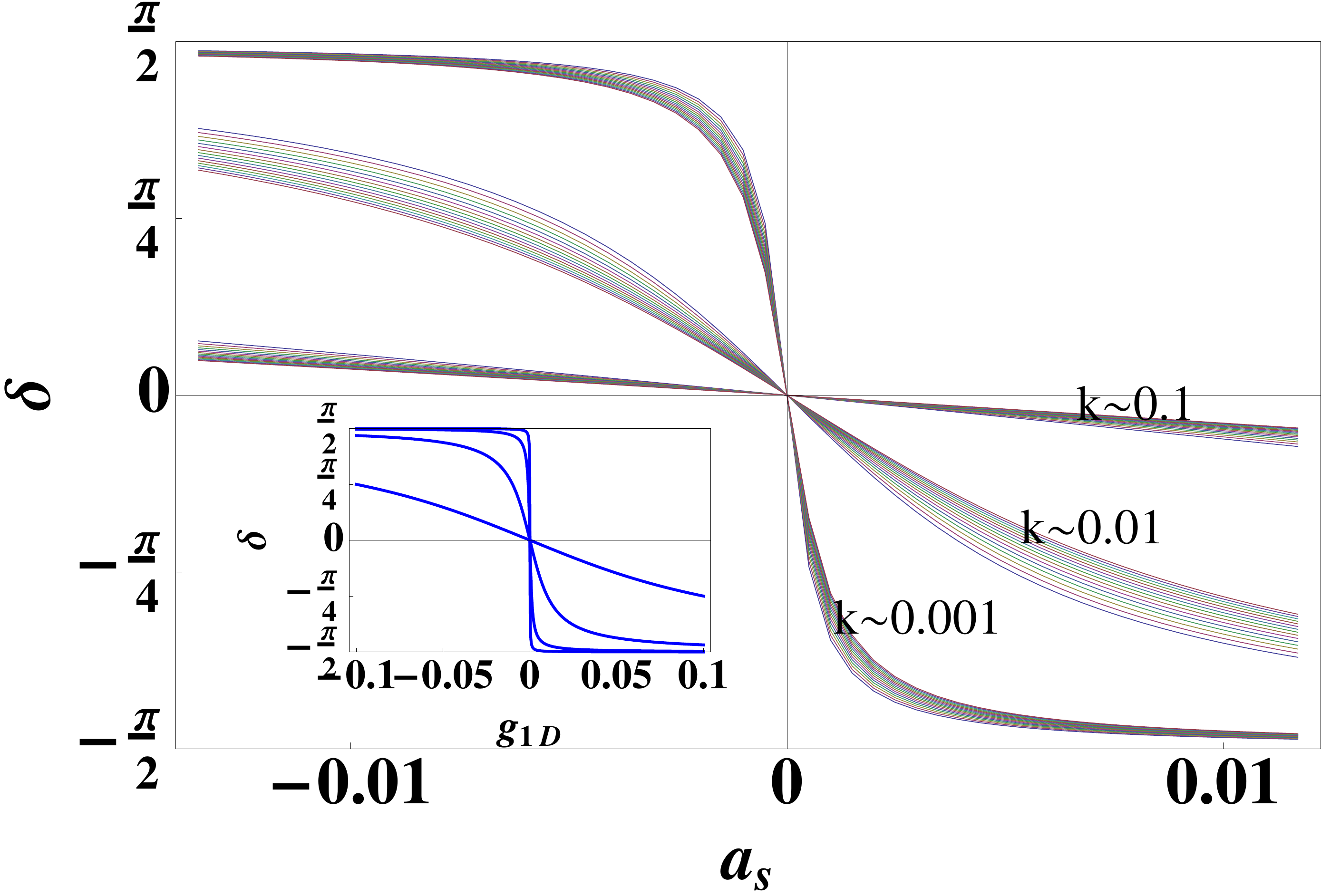}
\caption{Plot of the quasi-1D scattering phase shift in the transverse square well trapped system, as a function of the 3D scattering length $a_{s}$. The effective 1D interaction strength in this quasi-1D system is $\frac{g_{1D}}{L_{\perp}E_{0}}=\pi\frac{4a_{s}/L_{\perp}}{1-\Lambda_{E}(\frac{1}{2}E_{0})4a_{s}/L_{\perp}}$. $\Lambda_{E}(\epsilon)$ is the regularized Epstein zeta function discussed in the last section. $E_{0}=\frac{\hbar^2}{2\mu}\frac{4\pi^2}{L_{\perp}^2}$. The strips of curves clustered together are calculations for a range of energies in different regions of $k$: $k\sim 0.1, 0.01, 0.001$. $k$ denotes the momentum in $z$-direction, and the scattering energy $E_{z}=\frac{k^2}{2\mu}$. The inset is the pure-1D scattering phase shift as a function of $g_{1D}$, which comes into the effective 1D Hamiltonian as $V_{int}=g_{1D}\delta(z)$.}
\label{Quasi1Ddelta}
\end{figure}

In addition, the quasi-1D scattering phase shift recovers the behavior of the pure-1D scattering phase shift at perturbative values of the 3D scattering length. Fig. \ref{Quasi1Ddelta} is our numerical calculation for the square well transverse trapped system. The perturbative region $|a_{s}| \ll L_{\perp}$ is far away from the CIR region, and thus the transverse degree of freedom becomes effectively ``frozen". 
\section{Bound state energy from the frame transformation method}
As is well known from quantum defect theory\cite{Greene1987, Greene1980}, the reaction matrix $K$ can be used to calculate bound state energy and wave function. A key question is whether the $K$ matrix obtained from the local frame transformation can adequately predict the bound state energies in the quasi-1D system. As we show in the following paragraph, the local frame transformation method can also be applied to such a bound state calculation.
\par
In a calculation of true bound states, all the channels are closed, whereby the wave function outside the range of the potential is:
\begin{equation}
\Phi=\sum_{i,j}A_{j}(F^{c}_{i}\delta_{ij}-K_{ij}G^{c}_{i}),
\end{equation}
in which $i$ denotes the transverse channel indices, $j$ denotes the separate linearly-independent solutions, c denotes closed channels, $F$ and $G$ are regular and irregular solutions, $A_j$s is the vector of linear expansion coefficients for the $j$th independent solution. Since a the bound state energy lies below every threshold, both $F$ and $G$ will contain exponential divergent components. After imposing the asymptotically-decaying boundary condition at infinity, we find the equation that must be satisfied by the $A_j$:
\begin{equation}
(i\underline{\mathit{I}}+\underline{K})A=0,
\label{KMatrixEquation}
\end{equation}
in which $\underline{\mathit{I}}$ is the identity matrix, and $A$ is $n\times 1$ column vector of $A_{j}$. In order for this to yield a non-trivial solution to this system of homogenous equations, the determinant of the matrix $(-iI-K_{cc})$'s must vanish. The closed-channel reaction matrix $K_{cc}^{1D}$ can be derived from the local frame transformation, as is shown above. So in the quasi-1D system with only one nonzero scattering phase shift, the determinant is equal to: 
\begin{equation}
\det|-i\mathit{I}-K_{cc}|=-i(1-i\tan\delta_{l}\sum_{i=\mbox{all channels}}U_{i0}^2)
\end{equation}
As an example, for the \textit{s}-wave bound state, the relation between the bound state energy ($E=E_{i}-\frac{\hbar^2k_{i}^2}{2\mu}$) and the scattering length is:
\begin{equation}
-\dfrac{1}{a_{s}}=\dfrac{\partial}{\partial z}(z 2\pi  \sum_{i}\phi_{i}^2(x=0,y=0)\dfrac{1}{N_{i}^2}\dfrac{e^{-k_{i}|z|}}{k_{i}})|_{z\to0^{+}},
\end{equation}
in which $N_{i}$ is the normalization constant of channel $i$ in the $xy$-plane, corresponds to channel wave function $\phi_{i}$ and channel energy $E_{i}$. This results in a similarly divergent summation as we found in working out the quasi-1D scattering phase shift. Thus a similar regularization procedure is applied to find the energy of a weakly bound state as a function of the 3D scattering length. 
\par
In the case of 2D transverse isotropic harmonic confinement, we have compared the bound state energies with the analytical theory and confirmed that theses agree as well. We observe similar features of the finite range model potential correction to the zero-range result as was seen in \cite{PhysRevLett.91.163201}. 
\par
\section{Conclusion}
A general method has been formulated in this study whereby the local frame transformation method can be applied to a quasi-1D scattering process and to corresponding bound state problems. The formulation predicts the CIR position for a general transverse confinement. Our proposed procedure has been tested on a system with 2D square well confinement in the transverse directions, as well as for 2D oscillator transverse confinement where we agree with previous studies. The value of $\frac{L_{\perp}}{a_{s}(0)}=5.864(\pm 0.008)$ for the confinement-induced resonance with a hard square-well confinement in the transverse direction is, interestingly enough, very close to a factor of $4$ larger than the value of $\dfrac{L_{\perp}}{a_{s}(0)}=\dfrac{a_{ho}}{a_{s}(0)}=1.4603...$ that was obtained by Olshanii for the 2D transverse harmonic confinement. As far as we can tell, this simple relationship is merely a coincidence.
\par
It should be possible to extend this regularization method to general 2D confinement systems having various boundary conditions, and this direction is ripe for exploration in the future.
\section{Acknowledgement}
This work was supported in part by NSF. We thank Jose D'Incao, Yujun Wang, Doerte Blume, Senarath de Alwis, and Yuan Sun for insightful discussions.
\appendix
\section{Epstein zeta function}
The generalized form of the Epstein zeta function \cite{e1994zeta} in the first chapter is:
\begin{equation}
\begin{array}{lll}
Z\begin{bmatrix}
\overrightarrow{g} \\
\overrightarrow{h} \\
\end{bmatrix}^{\epsilon}(s)_{\phi}=
Z\begin{bmatrix}
g_1 & \cdots & g_{p}\\
h_{1} & \cdots & h_{p}\\
\end{bmatrix}^{\epsilon}(s)_{\phi}\\
\\
={\sum_{m_{1}, \cdots, m_{p}=-\infty}^{\infty}}^{\prime}[\psi(\overrightarrow{m}+\overrightarrow{g})+\epsilon]^{-s/2}e^{2\pi i(\overrightarrow{m}, \overrightarrow{h})},
\end{array}
\end{equation}
in which $p$ is a positive integer, $\overrightarrow{g}$ and $\overrightarrow{h}$ are $p$-dimensional real vectors, $g_{i}, h_{i}\in \it{R}$, $\overrightarrow{m}$ is a $p$-dimensional integer vector, $m_{i}\in \it{Z}$. $\phi(x)$ is quadratic form of vector $x$, $\phi(x)=\sum_{\mu,\nu}^{p} c_{\mu\nu} x_{\mu}x_{\nu}$, in which $c$ is $p\times p$ non-singlular symmetric matrix associated with $\psi$. The salar product of $p$-dimensional vectors $(\overrightarrow{g},\overrightarrow{h})=\sum_{\nu=1}^{p}g_{\nu}h_{\nu}$. $\epsilon$ is the ``inhomogeneity" of the generalized Epstein zeta function. The type of inhomogenous Epstein zeta function we use in this paper is the special case when $s=1$, $p=2$, $\overrightarrow{g}=(-\frac{1}{2},-\frac{1}{2})$, $\overrightarrow{h}=0$, namely:
\begin{equation}
\begin{array}{lll}
Z\begin{bmatrix}
-\frac{1}{2} & -\frac{1}{2} \\
0 & 0
\end{bmatrix}^{\epsilon}(\frac{1}{2})_{\phi}=\\
\\
{\sum_{m_{1},m_{2}=-\infty}^{\infty}}^{\prime}((m_{1}+\frac{1}{2})^2+(m_{2}+\frac{1}{2})^2+\epsilon)^{-\frac{1}{2}},
\end{array}
\end{equation}
in which $\phi(x)=\sum_{\mu=1}^{2} x_{\mu}^2$ in our case.
\section{$\underline{U}$ matrix elements}
\subsection{2D harmonic confinement}
In this case, at the radius $r$ we do the frame transformation, the energy normalized even parity channel function in 2D geometry is:
\begin{equation}
\psi^{H}_{n,m,q}(\mathbf{r})=(2\pi)^{-\frac{1}{2}}e^{im\phi}J_{m}(\sqrt{k^2-q^2}\rho)(\pi q)^{-\frac{1}{2}}\cos(q z),
\end{equation}
The energy normalized spherical wave function is:
\begin{equation}
F_{lm,k}(\mathbf{r})=Y_{lm}(\theta,\phi)\sqrt{\dfrac{2k}{\pi }} j_{l}(kr)
\end{equation}
The $\underline{U}$-matrix elements we use is the special case for $m=0$, $l=0$. For the angle part of the integration, formula 7.333 in \cite{citeulike:4561567} is applied. 
\begin{equation}\begin{array}{lll}
&&U_{n,m=0,q;l=0,m=0,\epsilon}=\dfrac{\int d\Omega Y_{00}(\hat{\mathbf{r}})\psi_{n,m,q}^{H}(\mathbf{r})}{j_{0}(kr)\sqrt{2k/\pi}}\\
\\
&=&\sqrt{\dfrac{1}{ka_{\perp}}}\left(\dfrac{k^2}{2\hbar\omega_{\perp}}-(2n+1)\right)^{-\frac{1}{4}}.
\end{array}
\end{equation}
\subsection{2D square well confinement}
In this case, at the radius $r$ where the frame transformation is performed, the energy normalized even parity channel function is:
\begin{equation}
\psi_{n_x,n_y,q}=\dfrac{2}{L_{\perp}}\cos(\dfrac{(2n_x+1)\pi x}{L_{\perp}})\cos(\dfrac{(2n_y+1)\pi y}{L_{\perp}})(\pi q)^{-\frac{1}{2}}\cos(qz)
\end{equation}
We use the spherical expansion of plane wave to get the $\underline{U}$-matrix elements: 
\begin{equation}
\begin{array}{lll}
\exp(i \mathbf{k}\cdot \mathbf{r})=\sum_{lm} i^{l}4\pi
Y_{lm}^{*}(\mathbf{\hat{k}})Y_{lm}(\mathbf{\hat{r}})j_{l}(kr),\\
\\
\cos(k_x x)\cos(k_y y)\cos(q z)=\dfrac{1}{8}(e^{ik_x x+ik_y y+iq z}+e^{ik_x x-ik_y y+iq z}\\
\\
+e^{ik_x x+ik_y y-iq z}+e^{ik_x x-ik_y y-iq z}+e^{-ik_x x+ik_y y+iq z}+\\
\\
e^{-ik_x x-ik_y y+iq z}+e^{-ik_x x-ik_y y-iq z}+e^{-ik_x x+ik_y y-iq z})\\
\\
=\dfrac{1}{8}\sum_{j}e^{i\mathbf{k}_{j}\cdot \mathbf{r}}
\end{array}
\end{equation} 
The energy normalization delta function $\delta(E-E^{\prime})$ comes into the following integration as well, so the on-shell $\underline{U}$-matrix elements can be obtained by deviding the delta function too.
\begin{equation}
\begin{array}{lll}
&&U_{n_x, n_y, q; l=0,m=0,\epsilon}=\dfrac{\int d\Omega Y_{00}(\hat{\mathbf{r}})\psi_{n_x, n_y, q}(\mathbf{r})}{j_{0}(kr)\sqrt{2k/\pi}}\\
\\
&=&\dfrac{1}{4L_{\perp}}\dfrac{\sum_{j}\int  d\Omega Y_{00}(\hat{\mathbf{r}})(\mathbf{r})e^{i\mathbf{k}_{j}\cdot \mathbf{r}}}{j_{0}(kr)\sqrt{2k/\pi}}\\
\\
&=&\dfrac{2}{L_{\perp}}\sqrt{\dfrac{1}{2\pi kq_{n_x,n_y}}}\\
\\
&=&\sqrt{\dfrac{4}{kL_{\perp}}}\left(\dfrac{k^2L_{\perp}^2}{4\pi^2}-(n_x+\dfrac{1}{2})^2-(n_{y}+\dfrac{1}{2})^2\right)^{-\frac{1}{4}}
\end{array}
\end{equation}

\end{document}